\documentclass[aps,prb,nofotinbib,superscriptaddress,twocolumn]{revtex4}
\usepackage{graphicx}
\usepackage{bm}
\usepackage{ulem}
\usepackage{color}

\newcommand{\bnmr}{$\beta$-NMR}
\newcommand{\bnqr}{$\beta$-NQR}
\newcommand{\STO}{SrTiO$_3$}
\newcommand{\Li}{$^8$Li}
\newcommand{\Tcb}{$T_c^{\rm bulk}$}
\newcommand{\Lip}{$^8$Li$^+$}
\newcommand{\bvec}{($\frac{1}{2}$ $\frac{3}{2}$ $\frac{1}{2}$)}

\begin{document}
\title{Depth Dependence of the Structural Phase Transition of \STO\ Studied with \bnmr\ and Grazing 
Incidence X-ray Diffraction}

\author{Z.~Salman}
\email{zaher.salman@psi.ch}
\affiliation{Clarendon Laboratory, Department of Physics, Oxford University, Parks Road, Oxford OX1 3PU, UK}
\affiliation{Laboratory for Muon Spin Spectroscopy, Paul Scherrer Institut, CH-5232 Villigen PSI, Switzerland}
\author{M.~Smadella}
\affiliation{Department of Physics and Astronomy, University of
  British Columbia, Vancouver, BC, Canada V6T 1Z1}
\author{W.~A.~MacFarlane}
\affiliation{Department of Chemistry, University of British Columbia,
  Vancouver, BC, Canada V6T 1Z1}
\author{B.~D.~Patterson}
\author{P.~R.~Willmott}
\affiliation{Swiss Light Source, Paul Scherrer Institut, CH-5232 Villigen PSI, Switzerland}
\author{K.~H.~Chow}
\affiliation{Department of Physics, University of Alberta, Edmonton,
  AB, Canada T6G 2J1}
\author{M.~D.~Hossain}
\author{H.~Saadaoui}
\author{D.~Wang}
\affiliation{Department of Physics and Astronomy, University of
  British Columbia, Vancouver, BC, Canada V6T 1Z1}
\author{R.~F.~Kiefl}
\affiliation{Department of Physics and Astronomy, University of
  British Columbia, Vancouver, BC, Canada V6T 1Z1}
\affiliation{TRIUMF, 4004 Wesbrook Mall, Vancouver, BC, Canada, V6T 2A3}

\begin{abstract}
  We present an investigation of the near-surface tetragonal phase
  transition in SrTiO$_3$, using the complementary techniques of
  beta-detected nuclear magnetic resonance and grazing-incidence X-ray
  diffraction. The results show a clear depth dependence of the phase
  transition on scales of a few microns.  The measurements support a
  model in which there are tetragonal domains forming in the sample at
  temperatures much higher than the bulk phase transition temperature.
  Moreover, we find that these domains tend to form at higher
  temperatures preferentially near the free surface of the crystal.
  The details of the tetragonal domain formation and their
  depth/lateral dependencies are discussed.
\end{abstract}
\maketitle

\section{Introduction}
All phase transitions in condensed matter, other than Bose-Einstein
condensation, arise because of interactions between the basic
constituents, e.g. spins, ions, electrons etc.  Near a surface or
interface, the symmetry of these interactions is broken and thus, in
general, the phases and phase transition properties (order parameter,
transition temperature etc.) will be altered
\cite{Binder74PRB,Mills71PRB,Pleimling04JPA}. In this paper, we
present an investigation of the effect of a free surface on the
well-known structural phase transition in SrTiO$_3$ (STO).  STO has a
number of interesting and useful properties. It is perhaps best known
for its use as a substrate for growing oxide thin films. More
recently, it was found that the interface between STO and other
insulating perovskites, such as LaAlO$_3$, may exhibit a variety of
unexpected properties, including 2-dimensional electric conductivity
\cite{Ohtomo04N,Thiel06S,Huijben06NM}, magnetism
\cite{Brinkman07NM,BenShalom09PRB} and even superconductivity at very
low temperatures \cite{Reyren07S}. Moreover, it was found that
superconductivity can be induced even at the free surface of STO
\cite{Ueno08NM}. All these unexpected properties make the surface and
near-surface region of STO of great fundamental interest as well as a
candidate for potential future applications.

Bulk STO undergoes a second-order structural phase transition at
$T_c^{\rm bulk} = 105$ K. The high-temperature phase is cubic, whereas
the low-temperature phase is characterized by a small tetragonal
distortion. The phase transition has been the subject of intense
experimental investigations
\cite{Cowley96PTSL,Ruett97EPL,Huennefeld02PRB,Osterman88JPC,Holt07PRL},
and its bulk properties are well understood. However, much less is
known about the behaviour close to a free surface or interface, in
particular its depth dependence
\cite{Ruett97EPL,Wang98PRL,Doi00PM,Mishina00PRL,Salman06PRL}. Here, we
report the use of two complementary techniques to better understand
the nature of the phase transition near the surface. As a reciprocal
space probe, grazing incidence X-ray diffraction (GIXRD) around the
critical angle is well suited to resolving changes in the average
periodic structure in the near-surface region of the
crystal\cite{Dosch92}. In addition, we probe the local structure in a
depth-resolved manner using low energy beta-detected nuclear magnetic
resonance (\bnmr).

These two techniques provide a unique picture of the structural phase
transition near the free surface. \bnmr\ measurements exhibit no depth
dependence up to $\sim 200$ nm from the surface. These results are
consistent with the X-ray measurements.  However, the latter extend
deeper into the sample, and we find that the phase transition varies
at depths of the order of a few $\mu$m.  From the results of both
local probe and scattering measurements, we conclude that static
domains of tetragonally distorted STO appear near the surface at
temperatures much higher than \Tcb ($\sim 50$K higher). The
temperature at which these domains appear depends on depth and varies
on a scale of a few $\mu$m.

\section{Experimental}
The \bnmr\ experiments were performed at the ISAC facility at TRIUMF.
In this part of the experiment, a beam of highly polarized \Lip\ is
implanted into the STO sample. Each implanted \Li\ decays (lifetime
$\tau=1.21$ s) emitting a $\beta$-electron preferentially opposite to
the direction of its polarization at the time of decay. Using
appropriately positioned detectors, one measures the asymmetry,
$A(t)$, of the $\beta^-$ decay along the initial polarization
direction ($z$) as a function of time, which is proportional to the
time evolution of the nuclear spin polarization, $P_z(t)$. The \Li\
implantation energy can be varied between $1-30$~keV, corresponding to
an average implantation depth of $1-200$~nm, allowing depth-resolved
\bnmr\ measurements. The GIXRD measurements were performed at the
surface diffraction station of the Materials Science beamline X04SA at
the SLS at the Paul Scherrer Institute. Here a beam of X-rays is
scattered from the STO single crystal, with different grazing incident
angles. By varying the incident angle, the mean scattering depth was
varied, allowing diffraction measurements at depths from a few nm up
to a few tens of $\mu$m.

The samples studied here are STO (100) single crystals supplied by
Crystal GmbH. Two crystals were used, in which the surface preparation
was slightly different. Sample {\bf 1} was studied as received, with a
polished surface, while sample {\bf 2} was etched using a HF buffered
solution and annealed at 950$^o$C in O$_2$ flux for 20 hours, to
ensure a flat, TiO$_2$ terminated surface \cite{Koster98APL}.

\section{Results}
Recently, we have demonstrated that zero-field spin relaxation and
beta-detected nuclear quadrupole resonance (\bnqr) of \Li\ can be used
as a probe of the near-surface transition behaviour in STO
\cite{Salman06PRL}. The quadrupole moment of the \Li\ nucleus couples
to the electric field gradients (EFG) of the crystal lattice.  Since
the EFG is sensitive to the local lattice geometry, the implanted \Li\
acts as an extremely sensitive probe of structural changes in the
lattice (e.g., due to a phase transition). The implanted \Li\ occupies
three equivalent sites in the STO lattice; the face-centered sites in
the Sr$^{2+}$-centered unit cell.\cite{MacFarlane03PB3}
\begin{figure}[h] \centering
  \includegraphics[width=0.9\columnwidth]{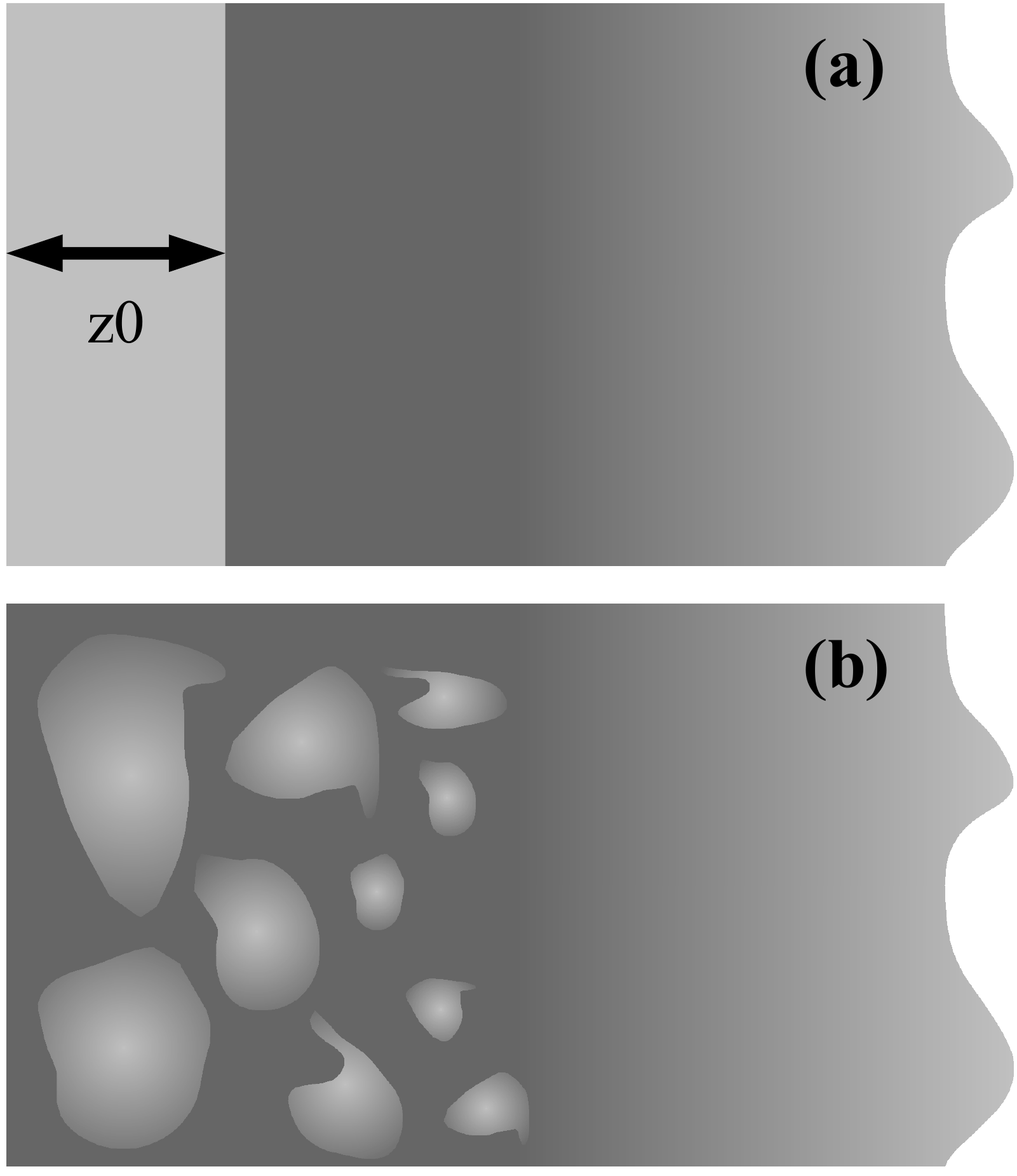}
  \caption{Illustration of the propagation of tetragonal domains from
    the surface of STO (left). The light and dark shaded areas
    represent tetragonal and cubic domains, respectively. (a) and (b)
    show scenario I and II, respectively.}
  \label{Scenarios}
\end{figure}
In the cubic phase (high temperatures), the EFG is axially symmetric,
with the main axis along Sr-\Li-Sr. At low temperatures, in the
tetragonal phase, the EFG in two of the three sites lose their axial
symmetry, depending on the direction of the tetragonal axis. In these
sites, the \Li\ spin polarization precesses rapidly, leading to an
immediate loss of the measured initial polarization.  This loss is a
direct indication of the onset of the tetragonal phase transition (at
least in a fraction of the sample volume). As expected, we observe a
loss of polarization below the phase transition. However, upon
cooling, this loss begins at temperatures as high as $\sim 150$ K,
indicating that the transition, at least in part of the probed volume
of the sample, occurs at a much higher temperature than \Tcb. These
results were attributed to an enhancement of the phase transition
temperature near the surface of STO, and are in agreement with optical
measurements \cite{Mishina00PRL}.

The observed enhancement of the transition temperature near the
surface could be explained using one of two more general scenarios. In
scenario I the order parameter has a depth ($z$) dependence but no
lateral ($x,y$) dependence. This results in an apparent transition
temperature that depends on $z$, changing gradually from a surface
value to \Tcb.\cite{Pleimling04JPA} In this scenario we expect that
the transition starts at the surface and propagates continuously into
the bulk as $T$ is decreased. This is illustrated in
Fig.~\ref{Scenarios}(a), where at any given temperature a layer of
thickness $z_0$ is in the tetragonal phase with $z_0$ increasing as
the temperature is decreased and diverges at \Tcb.  Alternatively, in
scenario II the order parameter depends on the depth, $z$, as well as
the lateral dimensions, $x$ and $y$. In this scenario tetragonal order
above \Tcb\ forms inhomogeneously in domains (possibly nucleating at
crystal defects \cite{Wang98PRL,Holt07PRL}), such that the average
volume fraction of these domains at any given temperature depends on
depth, as illustrated in Fig.\ref{Scenarios}(b). As we show below, a
measurement of the depth and temperature dependencies of the
tetragonal volume fraction supports the second scenario.

\begin{figure}[th]
  \centering
  \includegraphics[width=0.9\columnwidth]{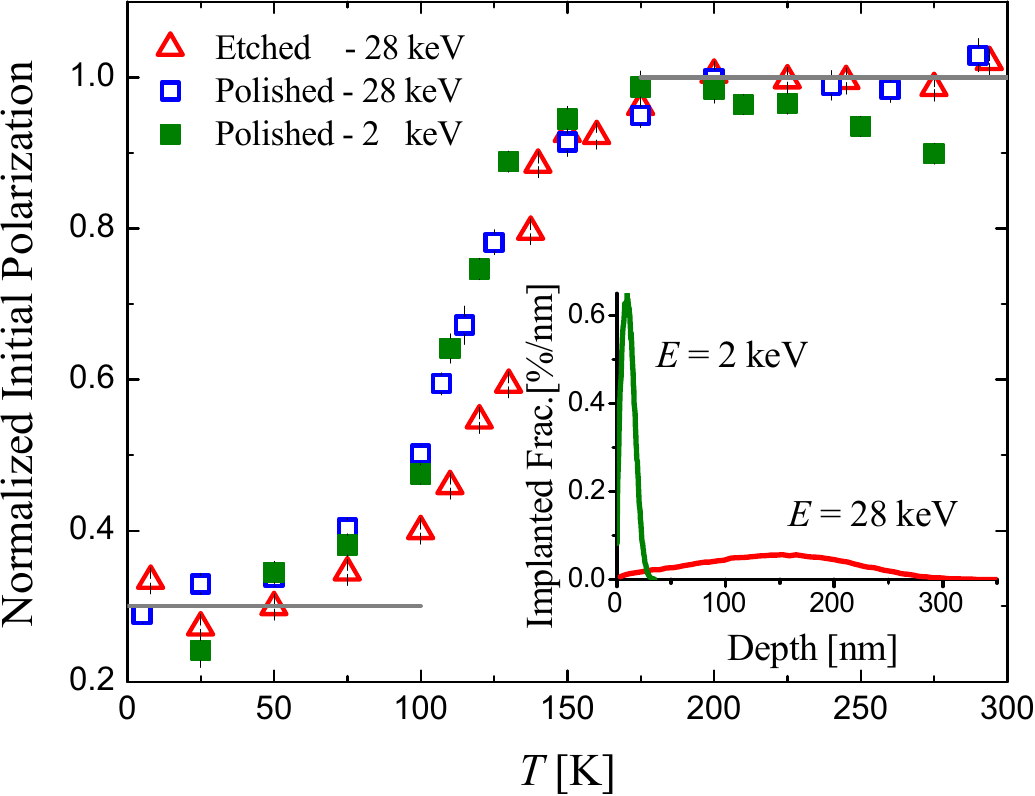}
  \caption{(Color online) The normalized initial polarization of \Li\
    as a function of temperature. The empty and filled squares are
    measurements on a polished STO sample and at implantation energy
    of 28 and 2 keV, respectively. The triangles are measurements on
    the etched sample at 28 keV. The inset shows the implantation
    profiles.}
  \label{AsyvsDepth}
\end{figure}
\begin{figure}[h]
  \centering
  \includegraphics[width=0.9\columnwidth]{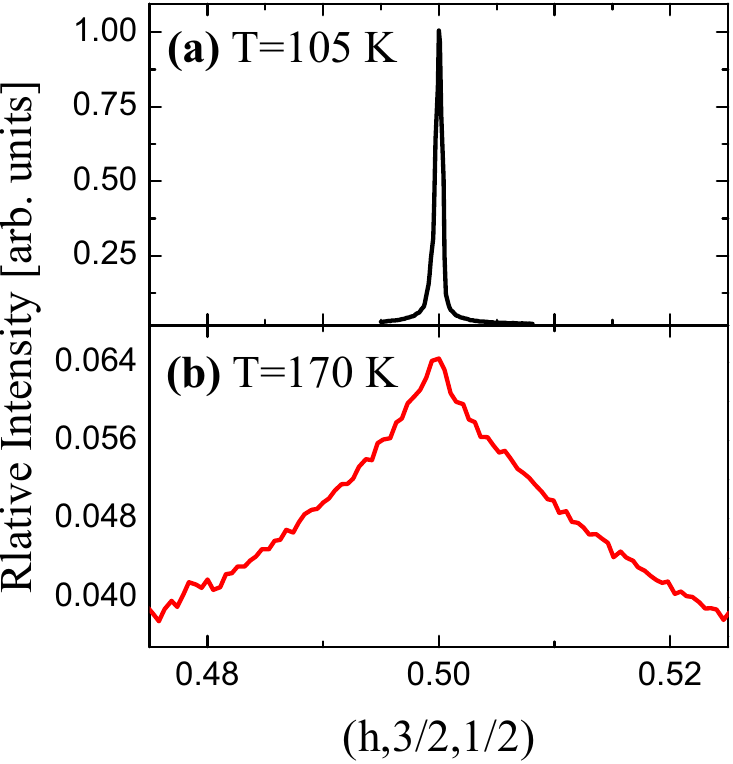}
  \caption{(Color online) The \bvec\ X-ray Bragg peak of the
    tetragonal structure measured at (a) T=105 K and (b) 170 K at the
    critical grazing incident angle for total external reflection of
    $0.15^o$.}
  \label{HighLowT}
\end{figure}
\begin{figure*}[hbt]
  \centering
  \includegraphics[width=1.5\columnwidth]{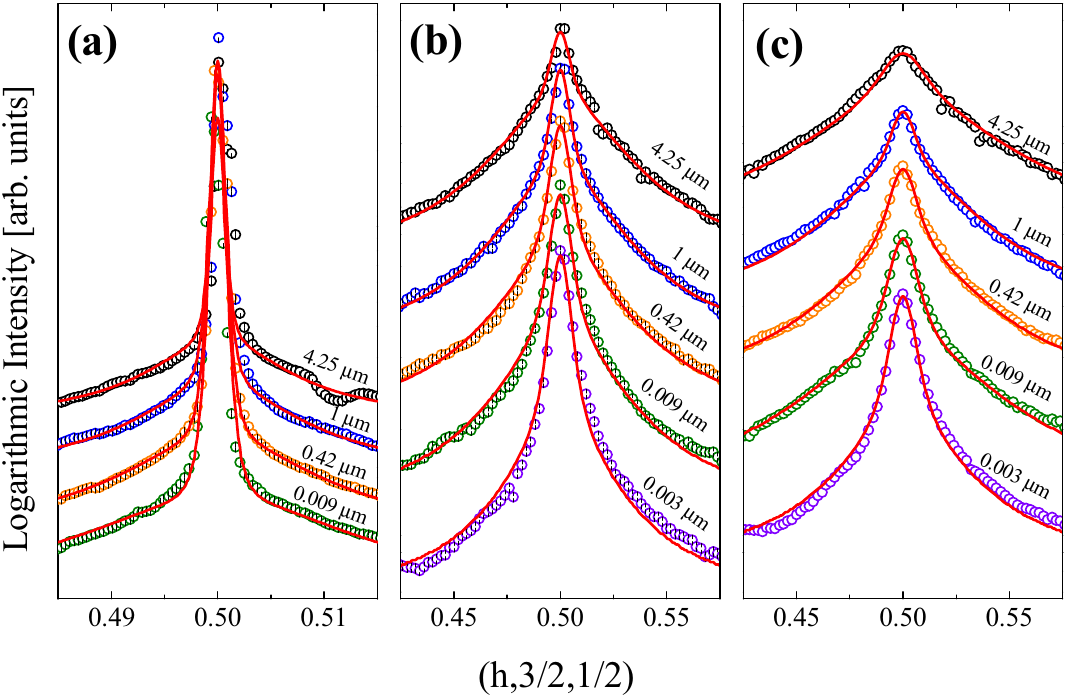}
  \caption{(Color online) The \bvec\ X-ray Bragg peak of the
    tetragonal structure measured at (a) T=120 K, (b) 130 K and (c)
    140 K at different grazing incident angles, corresponding to the
    quoted mean depth. The lines are the fits described in the text.}
  \label{LinesvsDepth}
\end{figure*}
A local probe such as \bnmr\ does not provide information on long
range order but is very sensitive to changes in the local point
symmetry. Also due to the size of the beam, \bnmr\ cannot provide
information regarding the lateral dimensions. However, it does enable
a depth-resolved measurement, which could, in principle, provide the
depth dependence of the order parameter averaged over lateral
directions $(x,y)$.  In order to try and identify which of the above
scenarios provides the best description, we performed a measurement of
the \Li\ polarization as a function of temperature at different
implantation depths. In scenario I, we expect the onset of
polarization loss to depend on depth. Figure~\ref{AsyvsDepth} shows
the initial \Li\ polarization as a function of temperature at two
implantation energies in both samples. We note first that the surface
preparation does not affect the onset of polarization loss
substantially.  In both etched and polished samples, and at an
implantation energy of 28 keV, the onset occurs at $\sim 150$ K.
Nevertheless, it is much sharper in the etched sample, possibly due to
the smoother surface and/or reduced defect concentration due to the
etching/annealing treatment. Additionally, we do not observe any
significant dependence on the \Li\ implantation depth in the range
10-200 nm. Although these results do not rule out either of the
scenarios, they allow us to put a lower limit on the scale of the
depth dependence of this effect, i.e., it must be larger than $\sim
200$ nm. Unfortunately, the limited energy of the implanted \Li\ does
not allow one to probe deeper into STO \cite{Smadella09PB}.

\begin{figure}[bh]
  \centering
  \includegraphics[width=0.9\columnwidth]{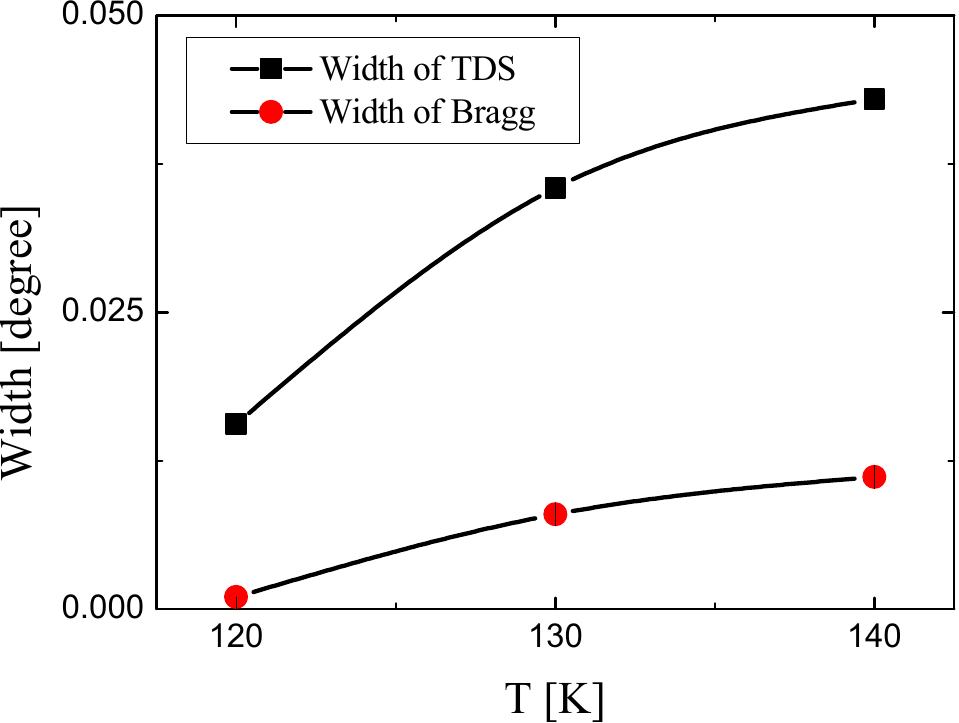}
  \caption{(Color online) The values of the width of the TDS and Bragg
    component as a function of temperature from fits to
    Eq.~(\ref{LpL2}). The solid lines are guide to the eye.}
  \label{WidthvsT}
\end{figure}

Information regarding the lateral, as well as the depth, dependence of
the phase transition on scales larger than that of \bnmr\ can be
obtained using GIXRD. In these experiments, we measure the intensity
of the superlattice reflections around $(h,k,l)=$\bvec. This
reflection is absent in the cubic phase and therefore can be used to
monitor the development of the tetragonal phase. The GIXRD
measurements were performed using 16 keV X-rays at different
temperatures, and at each temperature with various incident angles
relative to the crystal surface (accurate to $0.002^o$)
\cite{Willmott05ASS}. For STO, the critical grazing incident angle for
total external reflection at 16 keV is $0.15^o$. At low temperatures
($\le 105$ K) we observed a very sharp diffraction peak
[Fig.~\ref{HighLowT}(a)], which we attribute to Bragg diffraction from
the tetragonal phase. In contrast, at high temperatures ($>150$ K),
only a broad component is observed, since the \bvec\ Bragg scattering
is forbidden in the cubic phase. This component is attributed to
lattice dynamics, i.e.  scattering by thermally populated phonons or
thermal diffuse scattering (TDS) from the full volume of the
sample\cite{Holt99PRL,Wang00PRB,Holt07PRL,NielsenXRay}. The
diffraction peaks at intermediate temperatures are shown in
Fig.~\ref{LinesvsDepth}; they consists of two components in this
temperature range: a narrow and a broad component. As the temperature
is decreased, the narrow component grows and eventually dominates the
low-temperature diffraction. All these peaks could be fit
satisfactorily with a broad Lorentzian plus a narrow squared
Lorentzian function
\cite{Yoshizawa82PRL,Ruett97EPL,Papoular97PRB,Holt07PRL},
\begin{equation} \label{LpL2}
I(h)=A_{\rm TDS} {\cal L}(w_{\rm TDS},h)+A_{\rm Bragg} {\cal L}^2(w_{\rm Bragg},h),
\end{equation}
where $w_{\rm TDS}/A_{\rm TDS}$ and $w_{\rm Bragg}/A_{\rm Bragg}$ are
the width/areas of the corresponding component. ${\cal L}(w,h)$
represents a Lorentzian function of $h$ with a width $w$. In these
fits we assume that, for a fixed temperature, the width of the TDS and
Bragg contributions are depth independent. Note that this assumption
may be a simplification, however, it give a good fit to the
experimental data and minimizes the number of free parameters.  The
extracted values of the width as a function of temperature are shown
in Fig.~\ref{WidthvsT}.
Below we discuss the relation between $A_{\rm TDS}$ and $A_{\rm
  Bragg}$ as a function of temperature and depth.

\section{Discussion}
\begin{figure}[h]
  \centering
  \includegraphics[width=0.9\columnwidth]{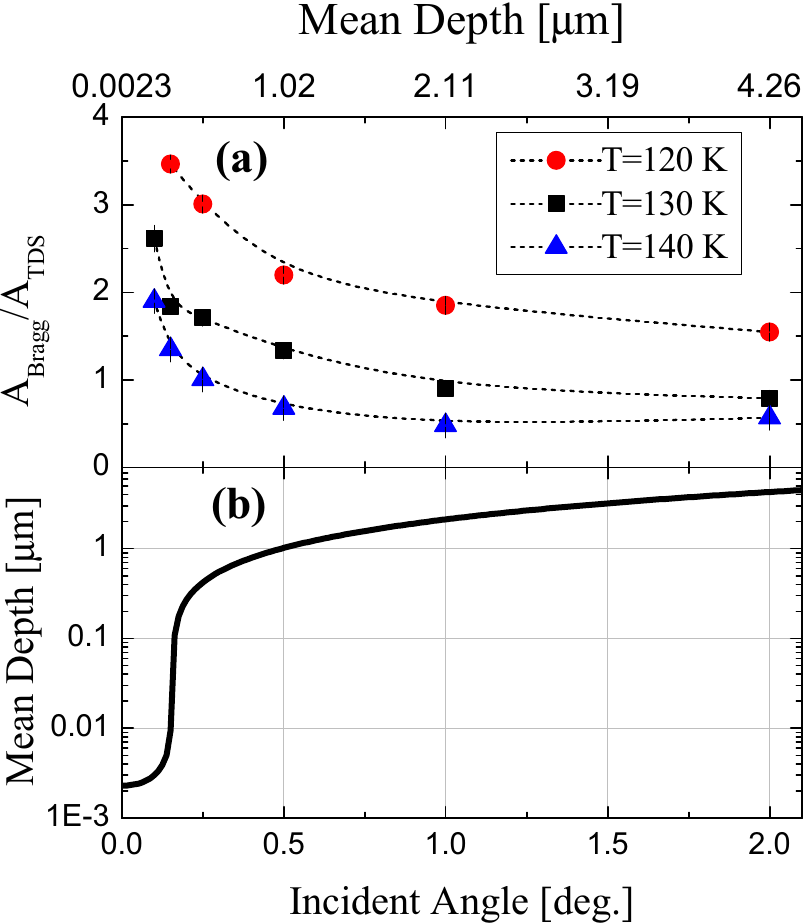}
  \caption{(Color online) (a) $A_{\rm Bragg}/A_{\rm TDS}$ as a
    function of incident angle or mean depth. The values for $T=120$ K
    were divided by a factor 15 to enable plotting all three
    temperatures on the same figure clearly. The dashed lines are a
    guide to the eye. (b) The calculated mean depth as a function of
    X-ray incident angle for STO with X-rays of energy 16 keV
    \cite{Henke93,XrayAttn}.}
  \label{ARatio}
\end{figure}

We start by looking at the area of the two components shown in
Fig.~\ref{LinesvsDepth} as a function of temperature. In
Fig.~\ref{ARatio}(a), we plot the ratio of the areas of the narrow to
broad component ($A_{\rm Bragg}/A_{\rm TDS}$) as a function of the
incidence angle at different temperatures. It clearly shows that the
contribution of the narrow component is larger at lower temperatures
and extends deeper into the sample. Note the narrow component is
present even at temperatures much higher than the bulk transition (see
Fig.~\ref{LinesvsDepth}).  This confirms that at least part of the
sample is non-cubic even above \Tcb, in agreement with the \bnmr\
measurements.

Next we discuss the dependence of the Bragg peaks on incidence angle.
As shown in Fig.~\ref{LinesvsDepth}, at all temperatures we find that
the relative intensity of the narrow component grows gradually as the
incident angle decreases, indicating that the contribution from the
non-cubic phase is enhanced in the near-surface region
[Fig.~\ref{ARatio}(a)].
The observed depth dependence of $A_{\rm Bragg}/A_{\rm TDS}$ clearly
exhibits that there is a ``crossover'' in the properties of the phase
transition from the bulk to the surface, which propagates smoothly and
gradually on a depth scale of a few microns, much longer than can be
detected with \bnmr. Moreover, we observe contributions from both the
tetragonal, as well as the cubic, phase at all measured depths and
intermediate temperatures.

Assume first a near-surface transition is described according to
scenario I [Fig.~\ref{Scenarios}(a)]. Then the contribution of the
narrow component is proportional to the tetragonal volume seen by the
X-ray beam,
\begin{equation} \label{Vtet}
A_{\rm Bragg} \propto S \int_{0}^{\infty}\Theta(z-z_{0})e^{-z/d}dz= Sd\left(1-e^{-z_0/d}\right),
\end{equation}
where $z_{0}$ is the thickness of the tetragonal layer at a given
temperature, $d$ is the mean penetration depth of the X-rays, $S$ is
the cross section area of the beam, and we assume, for simplicity, a
sharp cutoff on the tetragonal surface phase at $z_0$, i.e.,
$\Theta(z-z_{0})$ is the step function,
\begin{equation}
\Theta(z-z_{0})=\left\{ 
\begin{array}{ll}
  1 & z\leq z_{0}\\
  0 & z>z_{0}
\end{array}
\right. .
\end{equation}
Since the full volume of the sample contributes to the TDS at a given
temperature, we can write
\begin{equation}
  A_{\rm TDS} \propto S \int_{0}^{\infty} e^{-z/d}dz = Sd,
\end{equation}
Therefore the ratio of both contributions at a fixed temperature is
\begin{equation}
  \frac{A_{\rm Bragg}}{A_{\rm TDS}} \propto 1-e^{-z_0/d}.
\end{equation}
Note, this depth dependence causes this ratio to saturate when $d \ll
z_0$. Nevertheless, while $z_0 \sim 1$ $\mu$m, there is no evidence of
such behaviour for small mean depths. Here, we assumed an extreme case
with a step function, i.e. a certain depth value ($z_0$) separating
the tetragonal and cubic domains. A more gradual function with a
smoother separation could potentially show no leveling off for $d \ll
z_0$.  However, given the scale of $z_0$ from the scattering results,
we expect according to scenario I a very sharp loss of asymmetry as a
function of temperature in the \bnmr\ measurements, which is in
contrast with the results shown in Fig.~\ref{AsyvsDepth}. Therefore,
we conclude the results from both \bnmr\ and GIXRD techniques together
rule out scenario I and support scenario II instead. Unfortunately,
neither \bnmr\ nor GIXRD measurements provide clear information on the
lateral size/distribution of the tetragonal domains. It should be
pointed out here that the difference between tetragonal domains in
bulk and near the surface has been studied
experimentally\cite{Buckley99JAP,Chrosch98JPCM} and
theoretically\cite{Novak98JPCM}. These studies show a clear difference
between the two regions, in agreement with our results.

\section{Summary and Conclusions}
In conclusion, by combining the results of the \bnmr\ and GIXRD
measurements we have established a surprisingly long length scale
($\sim 1$ $\mu$m) for the depth dependence of the order parameter of
the phase transition in the near-surface region of STO. These results
confirm the findings of previous studies
\cite{Osterman88JPC,Doi00PM,Mishina00PRL,Salman06PRL}, which point to
variations in the nature of the structural phase transition in this
region. The results indicate that the phase transition in STO is
initiated locally within domains scattered through the volume of the
sample. The appearance of these domains depends on both temperature
and surface proximity. There is a strong tendency for these domains to
appear in near-surface regions even at temperatures much higher than
the bulk transition temperature (as high as $\sim 150$ K). The surface
alone cannot explain these observations, since it is not expected that
such effect will propagate more than a few monolayers into the sample
\cite{Binder74PRB,Mills71PRB,Pleimling04JPA}. This suggests defects
and strain near the surface region play an important role in the
nucleation of tetragonal domains \cite{Wang98PRL,Huennefeld02PRB}.
Finally, we note that the observed surface proximity effects reported
here may not be limited to the structural phase transition; it is
likely they may also contribute to the other intriguing phenomena
observed recently near the free surface of STO \cite{Ueno08NM} and its
interface with other insulating oxides
\cite{Ohtomo04N,Thiel06S,Huijben06NM,Brinkman07NM,BenShalom09PRB}.

\begin{acknowledgments}
  This research was supported by the Center for Materials and
  Molecular Research at TRIUMF, NSERC of Canada and the CIAR. Part of
  this work was performed at the Swiss Light Source, Paul Scherrer
  Institut, Villigen, Switzerland. We would like to acknowledge
  Michael Lange, Rahim Abasalti, Bassam Hitti, Donald Arseneau,
  Suzannah Daviel, Phil Levy and Matthew Pearson for expert technical
  support.
\end{acknowledgments}

\newcommand{\noopsort}[1]{} \newcommand{\printfirst}[2]{#1}
  \newcommand{\singleletter}[1]{#1} \newcommand{\switchargs}[2]{#2#1}

\end{document}